\documentclass[aps,prl,amsmath,amssymb,floatfix,twocolumn]{revtex4-1}
\usepackage{graphicx}
\usepackage{subfigure}
\usepackage{amsmath}
\usepackage{bbold}
\usepackage{slashed}
\usepackage{amssymb}
\usepackage{braket}
\usepackage{array}
\usepackage[colorlinks]{hyperref}
\usepackage{float}
\usepackage[margin=1in]{geometry}
\usepackage{times}
\usepackage{mathrsfs}

\newcommand{\ncmd}{\newcommand}
\ncmd{\nn}{\nonumber}
\ncmd{\pg}[1]{\textcolor{red}{#1}}
\ncmd{\mbf}[1]{\bs{#1}}
\ncmd{\Lam}{\Lambda}
\ncmd{\lam}{\lambda}
\ncmd{\Gam}{\Gamma}
\ncmd{\gam}{\gamma}
\ncmd{\sig}{\sigma}
\ncmd{\Dl}{\Delta}
\ncmd{\dl}{\delta}
\ncmd{\kap}{\kappa}
\ncmd{\Om}{\Omega}
\ncmd{\om}{\omega}
\ncmd{\mc}{\mathcal}
\ncmd{\eps}{\epsilon}
\ncmd{\veps}{\varepsilon}
\ncmd{\vphi}{\varphi}
\ncmd{\vtheta}{\vartheta}
\ncmd{\note}[1]{{\color{red}{#1}}}
\ncmd{\new}[1]{{\texttt{#1}  } }
\ncmd{\eq}[1]{Eq. \eqref{#1}}
\ncmd{\bs}{\boldsymbol}
\ncmd{\pll}{\parallel}
\ncmd{\dsty}{\displaystyle}

\begin{document}

\title{Fundamentals of crystalline Hopf insulators}
\author{Yuxin Wang$^{1}$}
\thanks{These authors contributed equally.}
\author{Alexander C. Tyner$^{2}$}
\thanks{These authors contributed equally.}
\author{Pallab Goswami$^{1,2}$}
\affiliation{$^{1}$ Department of Physics and Astronomy, Northwestern University, Evanston, IL 60208}
\affiliation{$^{2}$ Graduate Program in Applied Physics, Northwestern University, Evanston, IL 60208}

\date{\today}
\begin{abstract}
Three-dimensional, crystalline Hopf insulators are generic members of unitary Wigner-Dyson class, which can break all global discrete symmetries and point group symmetries. In the absence of first Chern number for any two-dimensional plane of Brillouin zone, the Hopf invariant $N_H \in \mathbb{Z}$. But in the presence of Chern number $N_H \in \mathbb{Z}_{2q}$, where $q$ is the greatest common divisor of Chern numbers for $xy$, $yz$, and $xz$ planes of Brillouin zone. How does $N_H$ affect topological quantization of isotropic, magneto-electric coefficient? We answer this question with calculations of Witten effect for a test, magnetic monopole. Furthermore, we construct $N$-band tight-binding models of Hopf insulators and demonstrate their topological stability against spectral flattening.  \end{abstract}
\maketitle

\underline{Introduction}: Beginning with the seminal work of Qi \emph{et al.}~\cite{ZhangFT}, many authors~\cite{EssinMagnetoelectric,Essin2010,malashevich2010theory,Coh2011,ryu2010,WangQiZhang,HughesProdan,FangGilbert,TurnerZhang,LuLee,chiu2016,AhnYang,LiSun} have argued for $\mathbb{Z}_2$-classification of isotropic magneto-electric (ME) coefficient $\frac{1}{3} \text{Tr}[\theta_{ij}] = 2 \pi \mathcal{CS}_{GS} $ of three-dimensional (3D) topological insulators (TIs), both in the presence and absence of time-reversal symmetry ($\mathcal{T}$). Here, $\theta_{ij}$ is the ME tensor and $\mathcal{CS}_{GS}$ is the Chern-Simons coefficient (CSC) of Berry connection of occupied bands. Arguments are based on the ambiguity of CSC under large non-Abelian gauge transformations of Bloch wave functions. Even TIs supporting $\mathbb{Z}$-classification of bulk invariant~\cite{ryu2010,LuLee} have been proposed to exhibit $\mathbb{Z}_2$-classification of  $\frac{1}{3} \text{Tr}[\theta_{ij}]$.

Can a 3D topological insulator (TI) support gauge-invariant CSC? In 2008, Moore \emph{et al.}~\cite{moore2008} showed that the CSC of $\mathcal{T}$-breaking, \emph{two-band Hopf insulators} (HIs) was gauge-invariant, and $\mathcal{CS}_{GS}=\frac{N_H}{2}$. Here, $N_H \in \mathbb{Z}$ is the Hopf invariant in the absence of Chern numbers for any two-dimensional (2D) plane of Brillouin zone (BZ)~\cite{pontrjagin1941}. This raises three important questions:  
\begin{enumerate}
\item Do HIs support $\mathbb{Z}_2$ or $\mathbb{Z}$ classification of $\frac{1}{3} \text{Tr}[\theta_{ij}]$ ?\label{q1}
\item Does $\mathbb{Z}_{2q}$-classified~\cite{pontrjagin1941} Hopf invariant of Hopf-Chern insulators (HCIs)~\cite{kennedy2016} cause topological quantization of $\frac{1}{3} \text{Tr}[\theta_{ij}]$?\label{q2} \item Are $N$-band HIs and HCIs stable against spectral flattening? \label{q3}
\end{enumerate}

These questions have remained unanswered, even though many authors have studied surface-states spectrum~\cite{deng2013, kennedy2016, liu2017}, bulk-boundary correspondence~\cite{alexandradinata2021,Zhu2021}, localization of Wannier charge centers over multiple unit cells~\cite{nelson2021, nelson2022,lapierre2021}, and possibilities of realizing HIs in engineered systems~\cite{ yuan2017, deng2018, schuster2019, unal2019, he2020, hu2020}. 
The primary goal of this Letter is to answer these questions with non-perturbative calculations of ME response in the presence of a test, magnetic monopole and explicit construction of $N$-band tight-binding models. We will critically address the stability of classifying space and gauge group of Berry connection against spectral flattening. 

\underline{2-band HIs and HCIs}: Two-band HIs, lacking $n$-fold rotational symmetries are described by a $2 \times 2$ Bloch Hamiltonian matrix
$H(\bs{k})=t[ n_0(\bs{k}) \sigma_0 + \bs{n}(\bs{k}) \cdot \boldsymbol \sigma ]$,
where the parameter $t$ has the dimension of energy, and 
\begin{eqnarray} \label{model1}&& n_1(\bs{k})=-2 u_0(\bs{k}) u_2(\bs{k}) + 2u_1(\bs{k}) u_3(\bs{k}), \nn \\
&& n_2(\bs{k})= 2 u_0(\bs{k}) u_1(\bs{k}) + 2u_2(\bs{k}) u_3(\bs{k}) , \nn \\
&& n_3(\bs{k})= u^2_0(\bs{k}) + u^2_3(\bs{k})-u^2_1(\bs{k})-u^2_2(\bs{k}),  \nn \\
&&  u_0(\bs{k})=(-\Delta+\sum_{i=1}^{3}\cos k_i ), 
 u_j(\bs{k})=\sin k_j + M_j, \nn \\
 \end{eqnarray} with $ j=1,2,3$. 
Here, $\Delta$ and $M_j$'s are dimensionless tuning parameters, and $|M_j|<1$. The lattice constant $a$ has been set to unity, and
\begin{equation}
 U(\bs{k}) = \frac{(u_0(\bs{k}) \sigma_0 + i \sum_{j=1}^{3} \; u_j(\bs{k}) \sigma_j) }{\sqrt{u^2_0(\bs{k}) +\sum_{j=1}^{3} u^2_j(\bs{k})}}  \end{equation} 
 is a suitable diagonalizing matrix of $H(\bs{k})$ that leads to non-singular Berry connection. 

\begin{figure}
\centering
\subfigure[]{
\includegraphics[scale=0.5]{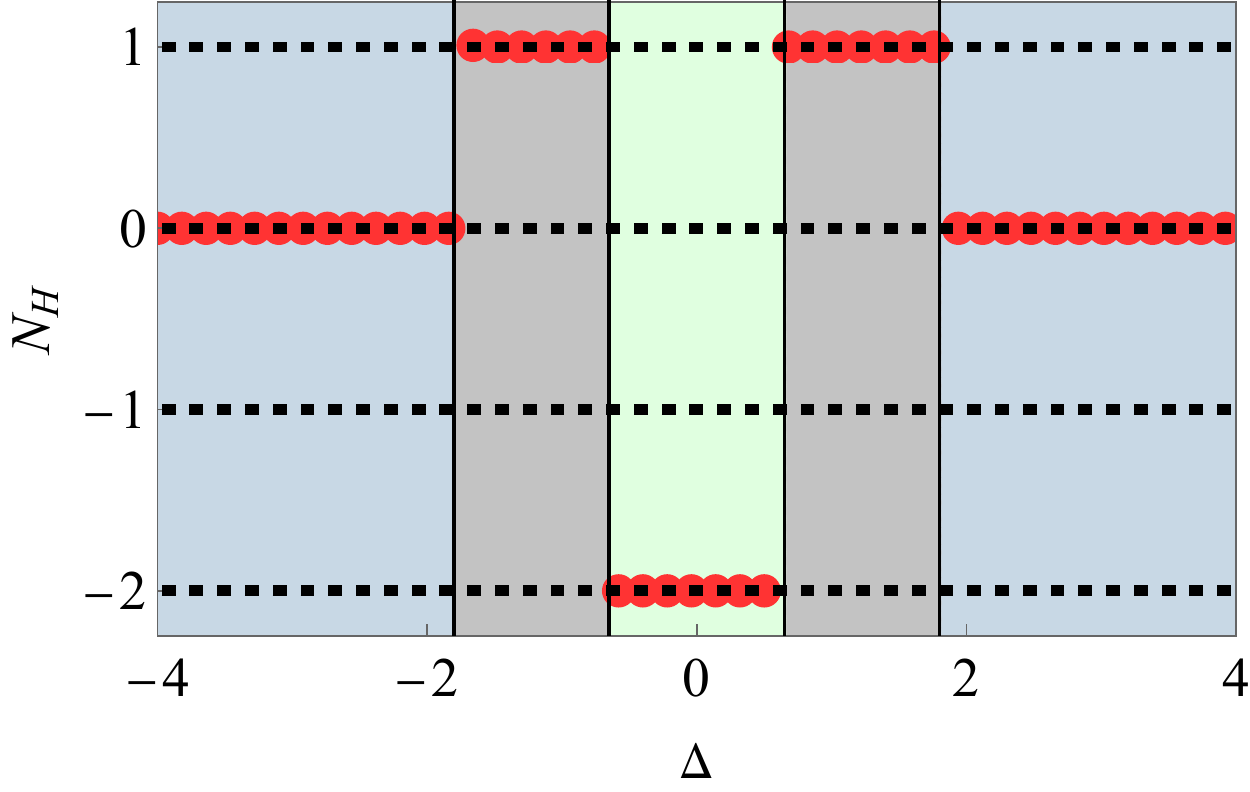}
\label{fig:BiML}}
\subfigure[]{
\includegraphics[scale=0.35]{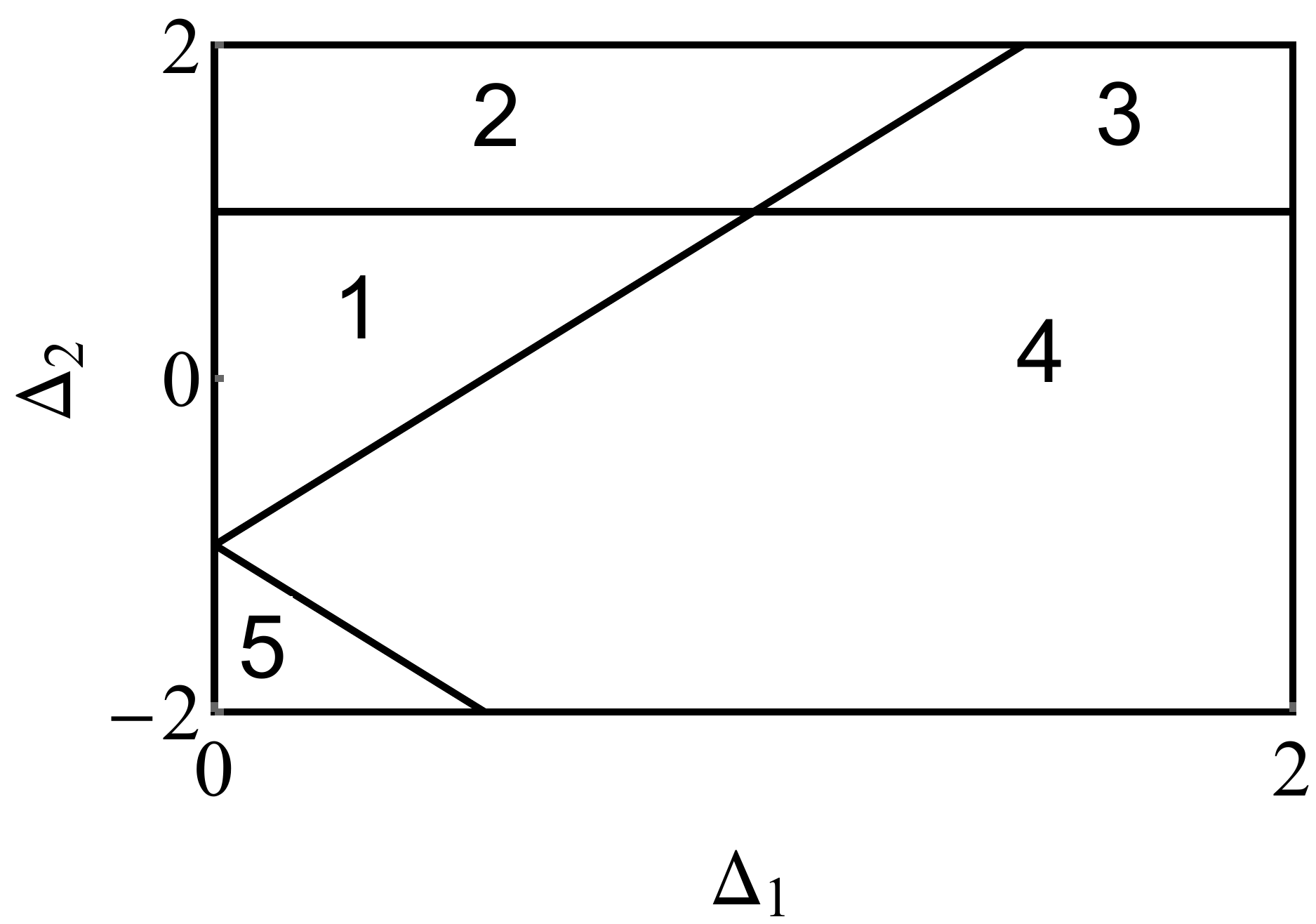}
\label{HopfChern}}

\caption{(a) Phase diagram of 2-band model of Eq.~\ref{model1} that supports $\mathbb{Z}$-classification of Hopf invariant under periodic boundary conditions, for $M_1=M_2=M_3=0.8$. (b) Phase diagram of 2-band Hopf-Chern insulator model of Eq.~\ref{HCI} for $\Delta_0=-1$ and the bulk invariants for phases 1, 2, 3, 4, and 5 are shown in Eq.~\ref{HCIinvariants}. }
\label{fig:PD}
\end{figure}

With $n_0(\bs{k}) \neq 0$, the model describes insulators, violating $\mathcal{T}$, charge-conjugation ($\mathcal{C}$), space-inversion ($\mathcal{P}$), $\mathcal{PT}$, $\mathcal{CP}$, $\mathcal{CT}$, and $\mathcal{CPT}$ symmetries. When $n_0(\bs{k})=0$, the particle-hole symmetry of conduction and valence bands corresponds to the $\mathcal{CP}$ symmetry: $\sigma_2 H^\ast(\bs{k}) \sigma_2 = -H(\bs{k})$. By setting $M_1=M_2=0$, one restores $\mathcal{C}_{4z}$-symmetry:
 \begin{equation}
 e^{i \frac{\pi}{4} \sigma_3} H(k_1,k_2,k_3) e^{-i \frac{\pi}{4} \sigma_3} = H(-k_2, k_1, k_3),
 \end{equation} 
and the special choice $M_j=0$ and $\Delta=\frac{3}{2}$ leads to the model of Ref.~\onlinecite{moore2008}. 

Since $U(\bs{k}) \in \frac{SU(2)}{U(1)} = S^2$  and $\bs{\hat{n}}(\bs{k})=\bs{n}(\bs{k})/|\bs{n}(\bs{k})|$ does not support 2D winding number for any plane of BZ (trivial second homotopy class), the third homotopy class of $\bs{\hat{n}}(\bs{k})$ is determined by the $\mathbb{Z}$-valued Hopf invariant 
 \begin{eqnarray}
 && N_H =  \frac{ \epsilon_{\mu \nu \lambda} }{24 \pi^2} \; \int_{T^3} d^3k  \; \text{Tr}[U^\dagger \partial_\mu U U^\dagger \partial_\nu U U^\dagger \partial_\lambda U]  \nn \\
&& =2\mathcal{CS}_{GS} = \frac{1}{4 \pi^2} \;   \int_{T^3} d^3k  \; \bs{A}(\bs{k}) \cdot \bs{B}(\bs{k}) ,
 \end{eqnarray}
 where $\bs{A}(\bs{k})= \frac{i}{2} \text{Tr}[\sigma_3 U^\dagger(\bs{k}) \bs{\nabla} U(\bs{k})]$, $\bs{B}(\bs{k})= \bs{\nabla} \times \bs{A}(\bs{k})$ are $U(1)$ Berry connection and curvatures of occupied valence band, respectively. 
The model of Eq.~\ref{model1} supports two distinct HIs with $N_H=+1$ and $N_H=-2$, as shown in Fig.~\ref{fig:PD}. When $M_j=0$, the phase diagram of $C_{4z}$-symmetric model is described by  
\begin{equation}
 N_H= -2  \Theta (1-|\Delta|) + \Theta (3-|\Delta|) \Theta (|\Delta|-1),
\end{equation} 
where $\Theta(x)$ is the Heaviside step function. By allowing longer range hopping terms in $u_0(\bs{k})$ we can obtain $N_H = \pm 1, \pm 2, \pm 3, \pm 4$~\cite{PRBTyner}.

\begin{figure*}[t]
\centering
\subfigure[]{
\includegraphics[scale=0.36]{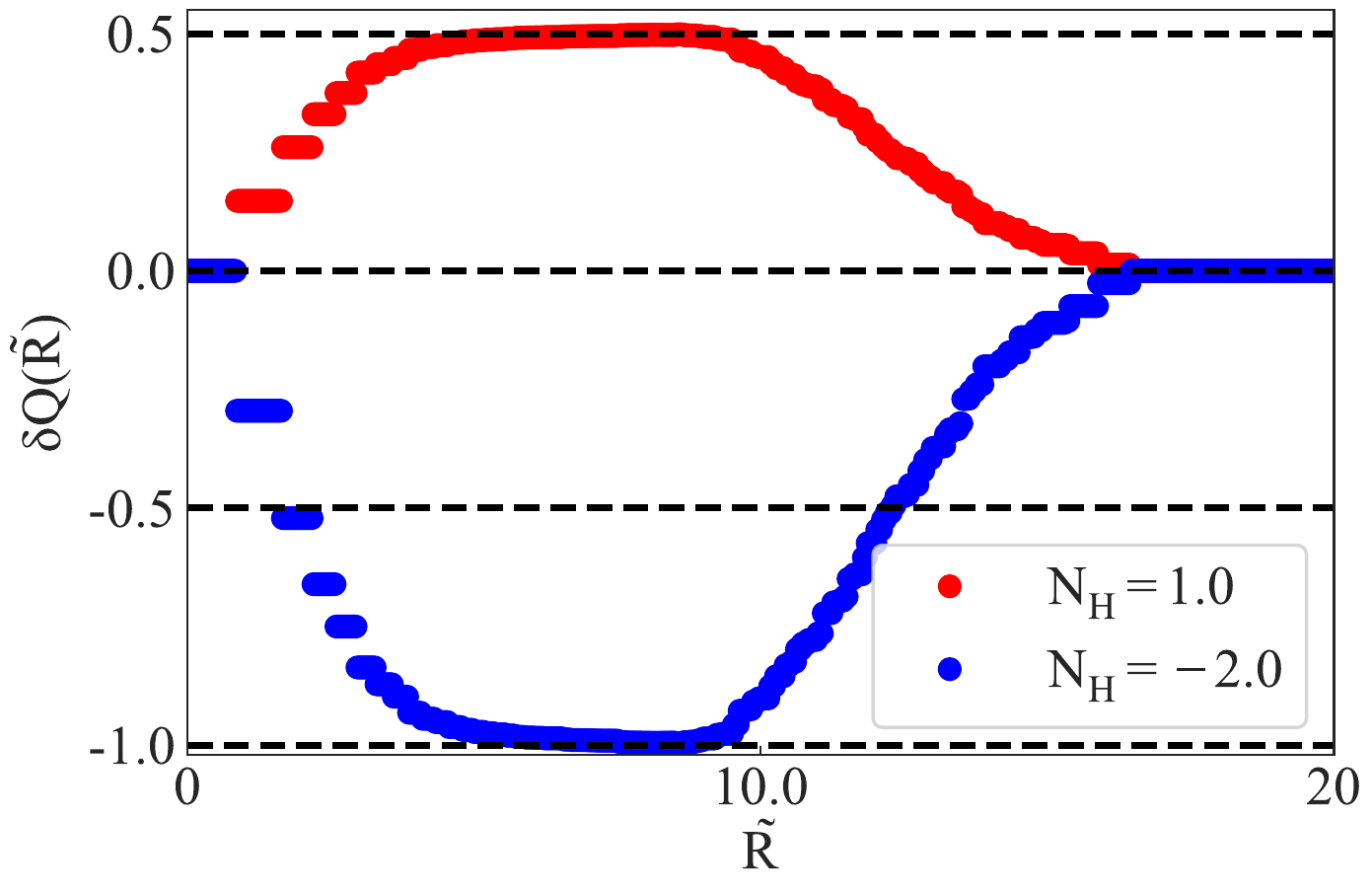}
\label{Witten4}}
\subfigure[]{
\includegraphics[scale=0.36]{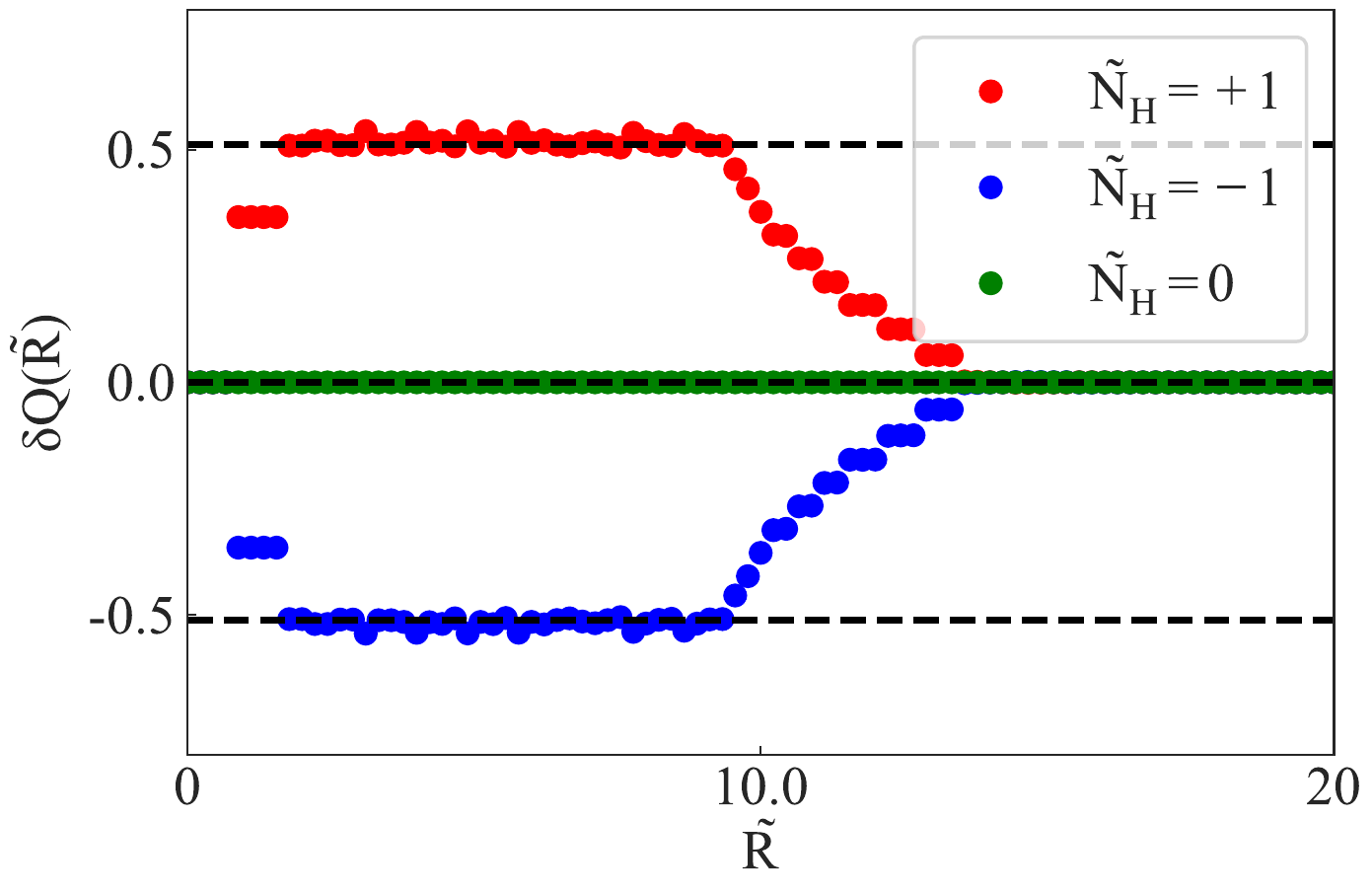}
\label{Witten6}}
\subfigure[]{
\includegraphics[scale=0.36]{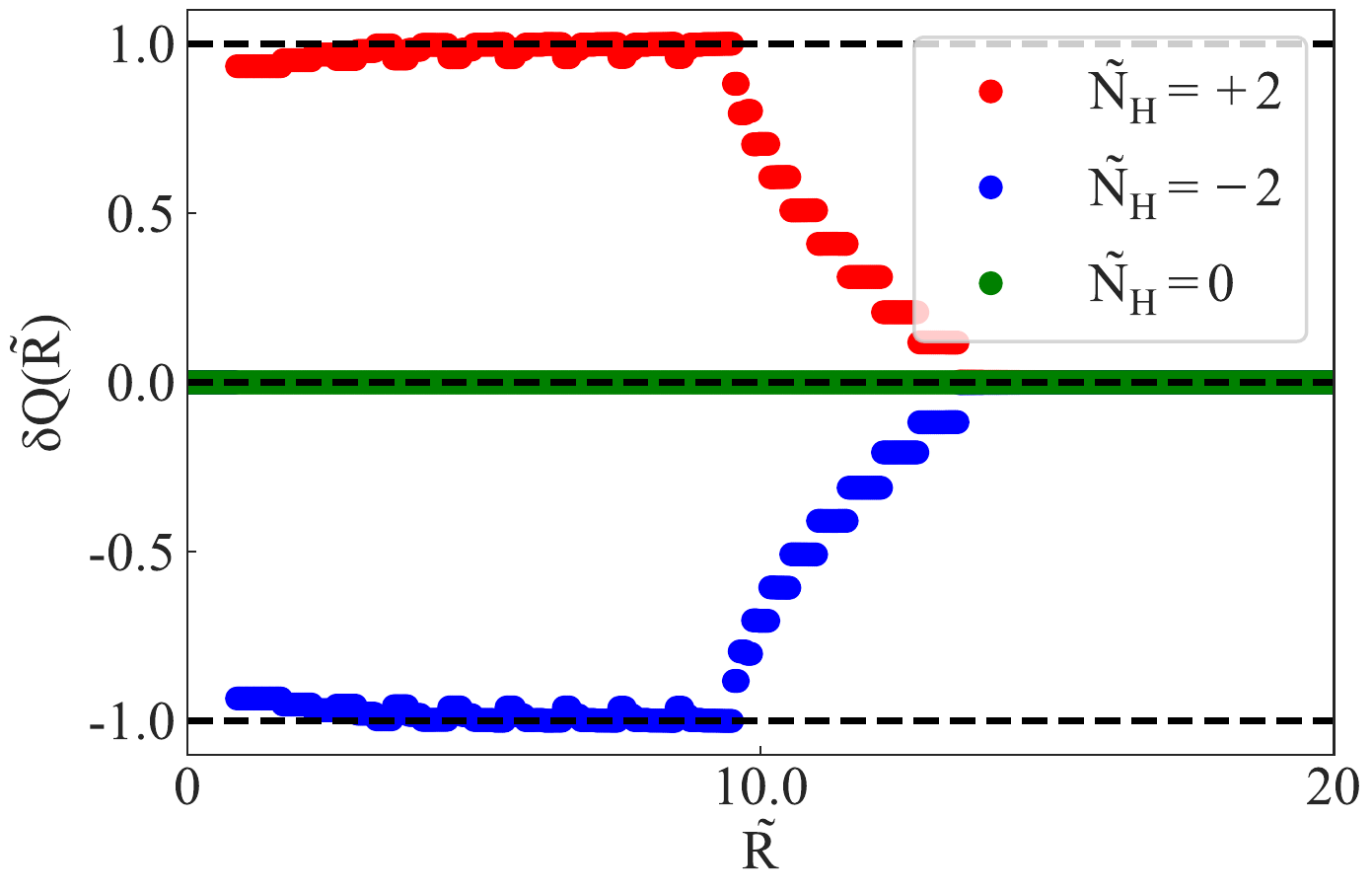}
\label{Witten5}}
\caption{Witten effect in the presence of a unit magnetic monopole. (a)  For Hopf insulators without Chern numbers, the maximum induced electric charge on monopole (in units of $-e$), for a system size $(\frac{L}{a})^3=20^3$ saturates to $\frac{N_H}{2}$ up to a numerical accuracy $10^{-4}$. The calculations for $N_H=1$ and $N_H=-2$ phases are performed for $(\Delta=2; M_j=0)$ and $(\Delta=0, M_j=0)$, respectively. (b)-(c) For Hopf-Chern insulators, the maximum induced electric charge on monopole detects signed invariant $\tilde{N}_H = |p| \mathfrak{C}_{xy}$, for $p = \pm 1$. Trivially stacked Chern insulators with $p=0$ do not support Witten effect.} 
\label{fig:witten}
\end{figure*}

In the presence of 2D winding number (first Chern number) 
\begin{equation}
\mathfrak{C}_{\mu \nu}=\frac{1}{4 \pi} \int dk_\mu dk_\nu \; \bs{\hat{n}}. (\partial_\mu \bs{\hat{n}} \times \partial_\nu \bs{\hat{n}}),
\end{equation}
$\mathbb{Z}_{2q}$-classification of $N_H$, where $q = \text{gcd}(\mathfrak{C}_{12}, \mathfrak{C}_{23}, \mathfrak{C}_{31})$, and its rigorous proof can be found in Refs.~\onlinecite{pontrjagin1941,Kapitansky1, Kapitansky2}. Simple models of HCIs can be constructed by considering twisted $O(3)$ skyrmion textures~\cite{deVega,Kundu,Nitta,kennedy2016}. We will work with 
\begin{eqnarray}\label{HCI}
 \begin{pmatrix} n_1(\bs{k}) \\ n_2(\bs{k}) \end{pmatrix} &=& \begin{pmatrix}\cos(p k_1) & -\sin (p k_1) \\ \sin(pk_1) & \cos(p k_1)  \end{pmatrix} \begin{pmatrix} \sin k_2 \\ \sin k_3 \end{pmatrix}, \nn \\
 n_3(\bs{k}) &=& -\Delta_0 + \Delta_1 (\cos k_2 + \cos k_3) \nn \\ && + \Delta_2 \cos k_2 \cos k_3,
\end{eqnarray}
possessing bulk invariants 
\begin{equation}
(N_H = p \times \mathfrak{C}_{23} \; \text{mod} \; 2 q ; \mathfrak{C}_{12}=0, \mathfrak{C}_{23}, \mathfrak{C}_{31}=0) , \\
\end{equation}
with $q= |\mathfrak{C}_{23}|$. A representative phase diagram is shown in Fig.~\ref{HopfChern}, for $p=\pm 1$, and various phases are distinguished by the following assignment of bulk invariants
\begin{eqnarray}\label{HCIinvariants}
&& \text{Phase 1:} (0; 0, 0, 0); \; \text{Phase 2:} (2 \; \text{mod} \; 4; 0, +2, 0); \nn \\ && \text{Phase 3:} (1 \; \text{mod} \; 2; 0, +1, 0); \text{Phase 4:} (1 \; \text{mod} \; 2; 0, -1, 0); \nn \\ &&\text{Phase 5:} (2 \; \text{mod} \; 4; 0, -2, -0).
\end{eqnarray}
For $p=0$, the phases correspond to trivially stacked Chern insulators, with $N_H=0$. 

\underline{Witten effect}: When a test, Dirac monopole~\cite{dirac1931quantised} with magnetic charge $g=\frac{\hbar m}{2e}$, and $m \in \mathbb{Z}$ is embedded in an \emph{infinite ME medium}, the monopole exhibits Witten effect~\cite{witten1979dyons} and binds electric charge
\begin{equation}\label{tracetheta}
\delta Q = \frac{-e m }{2 \pi} \; \frac{ \text{Tr}[\theta ]}{3} .
\end{equation}
To probe ME response of finite HIs and HCIs under open boundary conditions, we place a Dirac monopole at the center of the system $\bs{r}=(0,0,0)$, and use the singular vector potential
 \begin{eqnarray}
 \bs{a}(\bs{r}_i)= \frac{\hbar m}{2e} \; \frac{-z_i \hat{y} + y_i \hat{z}}{r_i(r_i+x_i)}, 
 \end{eqnarray}
that exhibits Dirac string singularity along the negative $x$ axis. The lattice sites are labeled by $\bs{r}_i=\frac{a}{2}(2n^x_i+1,2n^y_i+1,2n^z_i+1)$, with $n^a_i \in \mathbb{Z}$.
After obtaining a tight-binding model in real space by Fourier transformation of $H(\bs{k})$, the hopping parameters between different lattice sites $\bs{r}_i$ and $\bs{r}_j$ are multiplied by Peierls phase factors $e^{i\nu_{ij}}$, with $\nu_{ij}=-(e/\hbar)\int_{\bs{r}_i}^{\bs{r}_j}\mathbf{a} \cdot d\mathbf{l}$. The spectrum and eigenfunctions are solved in the presence and absence of monopole and we calculate the difference 
\begin{equation}\label{deltaQ}
\frac{\delta Q (\tilde{R})}{-e} = \sum_{\alpha \in occ} \sum_{|r_i| < R} (|\psi_\alpha(\bs{r}_i, m)|^2 - |\psi_\alpha(\bs{r}_i, m=0)|^2),
\end{equation} where $\tilde{R}=R/a$, and $R$ is the radius of a Gaussian sphere, centered at the monopole~\cite{rosenberg2010witten, tynermonopole}. The sum over $\alpha$ is performed for $(\frac{L}{a})^3$ occupied states to maintain half-filled condition. The results for 2-band HIs and HCIs are shown in Fig.~\ref{fig:witten}. For HIs without Chern numbers (see Fig.~\ref{Witten4} ), Witten effect detects $\mathbb{Z}$-classification of $\frac{1}{3} \text{Tr}[\theta_{ij}] = \pi N_H$. In contrast to this, Witten effect identifies $\frac{1}{3} \text{Tr}[\theta_{ij}]=\pi \tilde{N}_H$, with $\tilde{N}_H=|p| \mathfrak{C}_{23}$ for HCIs (see Figs.~\ref{Witten6} and \ref{Witten5} ). As $\frac{1}{3} \text{Tr}[\theta_{ij}]$ does not depend on the sign of $p$, Pontryagin's theorem for $\mathbb{Z}_{2q}$ classification is satisfied. But the isotropic ME coefficient does not follow $\mathbb{Z}_2$-classification.

\underline{Classifying space of $N$-band models}: A generic $N\times N$ Bloch Hamiltonian 
$H(\bs{k};N)= \sum_{j=1}^{N} E_j(\bs{k}) P_j(\bs{k})$
of class A insulators can be diagonalized by a unitary transformation
\begin{equation}\label{diagonalization}
 \mathscr{U}^\dagger (\bs{k}) H (\bs{k};N) \mathscr{U}(\bs{k}) 
=\mathrm{diag}[E_{N}(\bs{k}),...,E_1(\bs{k})], \end{equation}
where $E_j(\bs{k})$ is the energy eigenvalue and $P_j(\bs{k})$ is the projector of $j$-th band. We assume that the non-degenerate, energy eigenvalues can be organized in an ascending order, i.e., $E_{j+1}(\bs{k}) > E_j(\bs{k})$ over the entire BZ.  
As $H(\bs{k}; N)$ remains invariant under \emph{admissible} $[U(1)]^{N}$ gauge transformation 
\begin{equation}
\mathscr{U}(\bs{k}) \to \mathscr{U}(\bs{k}) (e^{-i \vartheta_N(\bs{k})} \oplus ... \oplus e^{-i \vartheta_1(\bs{k})}) \end{equation}
of band eigenfunctions, 
\begin{equation}
\mathscr{U}(\bs{k}) \in UFM(N)=\frac{U(N)}{[U(1)]^N}=\frac{SU(N)}{[U(1)]^{N-1}}.
\end{equation} This coset space is distinct from the complex Grassmanian manifold \begin{equation} Gr(N,M)=\frac{U(N)}{U(N-M) \times U(M)},\end{equation} used for K-theory analysis of class A insulators, after spectral flattening of $M$ occupied and $(N-M)$ unoccupied bands. Only for the special case of 2-band models, with $N=2$, $M=1$, $UFM(2) = Gr(2,1)=S^2$.  

Employing the projectors and canonical kets (orbital basis), we can define a general form of diagonalizing matrix
\begin{eqnarray}\label{gauge1}
&& \mathscr{U}(\bs{k}) = \left( |\psi_{N}(\bs{k}) \rangle,..., |\psi_{1}(\bs{k}) \rangle \right), \nn \\ &&= \left( \frac{P_{N}(\bs{k}) | N \rangle e^{-i\vartheta_N(\bs{k})}}{\sqrt{\langle N | P_{N}(\bs{k}) |N \rangle }},...,\frac{P_{1}(\bs{k}) | 1 \rangle e^{-i\vartheta_1(\bs{k})}}{\sqrt{\langle 1 | P_{1}(\bs{k}) |1 \rangle }} \right), \nn \\
\end{eqnarray}
The intra- and inter- band components of flat, $U(N)$ Berry connection matrix 
$A_\mu = -i \mathscr{U}^\dagger(\bs{k}) \partial_\mu \mathscr{U}(\bs{k})$
are given by
\begin{eqnarray}
A_{\mu, jj}(\bs{k}) &=& - i \langle \psi_j(\bs{k}) | \partial_\mu | \psi_j(\bs{k}) \rangle \nn \\ &=& -i \frac{\langle j | [P_j(\bs{k}), \partial_\mu P_j(\bs{k})] | j \rangle}{2 \langle j |P_j(\bs{k}) | j \rangle } -\partial_\mu \vartheta_j(\bs{k}), \label{gauge3} \\
 A_{\mu, jl}(\bs{k}) &=& - i \langle \psi_j(\bs{k}) | \partial_\mu | \psi_l(\bs{k}) \rangle \nn \\ &=&-i e^{i(\vartheta_j(\bs{k}) -\vartheta_l(\bs{k})) } \frac{\langle j | P_j(\bs{k}) \partial_\mu P_l(\bs{k}) | l \rangle}{\sqrt{ \langle j |P_j(\bs{k}) | j \rangle  \langle l |P_l(\bs{k}) | l \rangle} }. \label{gauge4} \nn \\
\end{eqnarray}
which respectively transform \emph{inhomogeneously and covariantly under admissible} $[U(1)]^N$ gauge transformation. 

While the Berry connection matrix of a ground state with $M$ occupied bands is defined by the $U(M)$ form 
\begin{equation}\label{Manybody}
A_{\mu,GS}(\bs{k})  = -i \begin{pmatrix} A_{\mu, MM}(\bs{k}) & \cdot & \cdot & \cdot &  A_{\mu, M1}(\bs{k})  \\
\cdot &  &  &  &  \cdot \\ 
\cdot &  &  &  &  \cdot \\ 
\cdot &  &  &  &  \cdot \\ 
A_{\mu, 1M}(\bs{k}) & \cdot & \cdot & \cdot & A_{\mu, 11}(\bs{k})  \end{pmatrix} ,
\end{equation}
it must be accompanied by the projected Bloch Hamiltonian 
$H_{GS} (\bs{k}) = \sum_{j=1}^{M} E_{j} (\bs{k}) P_ j (\bs{k})$
as a gauge-fixing operator. In any calculation for a model Hamiltonian, distinct energy eigenvalues remain fixed, as they are observables, and the gauge group of $A_{\mu,GS}(\bs{k})$ is fixed by $H_{GS} (\bs{k})$ to $ \frac{U(M)}{[U(1)^M]} $.

The second and third homotopy groups of unitary flag manifold $UFM(N)$ are given by
\begin{equation}\label{homotopyUFM}
 \pi_2(UFM(N))=\mathbb{Z}^{N-1}, \; 
 \pi_3(UFM(N)) = \mathbb{Z},
\end{equation}
which are important for constructing $SU(N)$ monopoles and skyrmion textures, respectively~\cite{carroll1993,shnir2006magnetic,Shifman}. Therefore, $UFM(N)$ allows $N$-band Hopf and Hopf-Chern insulators, which has been discussed in recent work by Lapierre \emph{et al.}~\cite{lapierre2021}. But the homotopy groups of complex Grassmanian manifold are given by
\begin{equation}
 \pi_2(Gr(N,M))=\mathbb{Z}, \; \pi_3(Gr(N,M)) = \delta_{N,2}\delta_{M,1} \mathbb{Z}. 
 \end{equation}
 Therefore, $N$-band HIs and HCIs identified from Eq.~\ref{homotopyUFM} appear to be unstable against spectral flattening. 
 
 To resolve this issue we carefully address subtleties of spectral flattening argument, which are generally overlooked by practitioners of $K$-theory. Let us define an adiabatic tuning parameter $t \in [0,1]$ and the spectral deformation~\cite{ZhangFT} as
\begin{eqnarray}
&& E_j(\bs{k},t) = E_j(\bs{k})(1-t) - t,  \; 1 \leq j \leq M,  \nn \\
&& E_j(\bs{k},t) = E_j(\bs{k}) (1-t) +  t, \; M+1 \leq j \leq N,  \nn \\ 
&& P_j(\bs{k}, t) =  P_j(\bs{k}, t=0) = P_j(\bs{k}), \nn \\
&& H(\bs{k}; N; t) =  \sum_{j=1}^{N} E_j(\bs{k}, t) P_j(\bs{k}).  \label{gauge5} \end{eqnarray}
Notice that $N$ mutually commuting projection operators are held fixed for any $t \in [0,1]$, which define a representation of Cartan sub-algebra for $\mathfrak{su}(N)$. When $t<1$, the band ordering remains unchanged and the simultaneous diagonalization of $P_j(\bs{k})$'s leads to $\mathscr{U}(\bs{k}, t<1) \in UFM(N)$. 

At $t=1$, the flattened Hamiltonian 
\begin{equation}
\mathcal{H}(\bs{k})=H(\bs{k}; N; t=1)= \sum_{j=M+1}^{N} P_j(\bs{k})  -  \sum_{j=1}^{M} P_j(\bs{k}), 
\end{equation}
displays $M$-fold and $(N-M)$-fold degeneracies of occupied and unoccupied bands, respectively. This apparently allows non-Abelian $U(M) \times U(N-M)$ gauge transformations of Bloch wave functions and the \emph{classifying space seems to change abruptly} from $UFM(N)$ to $Gr(N-M,M)$,
causing an \emph{abrupt change of homotopy groups}.

But $U(N-M) \times U(M)$ gauge transformations \emph{violate the assumption} in Eq.~\ref{gauge5} that $N$ \emph{mutually commuting projection operators remain fixed for all} $t \in [0,1]$. Hence, $U(N-M) \times U(M)$ gauge transformations are \emph{inadmissible}. The non-Abelian redundancy of basis is always removed by simultaneous diagonalization of $N$ commuting projectors, guaranteeing the stability of classifying space $UFM(N)$ and its homotopy groups. Therefore, $N$-band HIs and HCIs describe stable topological phases of matter.

 \underline{Representative $N$-band models}: 
To write $N$-band models of HIs, we have to embed a non-trivial $SU(2)$ matrix inside a $SU(N)$ matrix, such that $\mathscr{U}(\bs{k})$ describes $SU(N)$ skyrmion textures in momentum space, and
\begin{equation}\label{HIdef}
 W[\mathscr{U}] = 2 \sum_{j=1}^{N} \mathcal{CS}_j =  \sum_{j=1}^{N} \frac{1}{4 \pi^2}   \;  \int_{T^3} d^3k \; \bs{A}_{j} (\bs{k}) \cdot \bs{B}_{j} (\bs{k}).
\end{equation}
Here, $\bs{A}_j(\bs{k})$, $\bs{B}_j(\bs{k})=\bs{\nabla} \times \bs{A}_j(\bs{k})$, and $\mathcal{CS}_j$ respectively denote $U(1)$ Berry connection, curvatures, and the Chern-Simons coefficient of $j$-th band. By using gauge-fixed connection of Eq.~\ref{Manybody} one can verify that \begin{eqnarray}\label{A11}
&& \mathcal{CS}_{GS} =\frac{1}{8\pi^2} \;  \int d^{3}k  \; \epsilon_{\mu \nu \lambda} \; \text{Tr}[A_{\mu,GS}\partial_{\nu}A_{\lambda,GS} \nn \\ && +\frac{2i}{3} A_{\mu,GS}A_{\nu,GS}  A_{\lambda,GS}]
=\sum_{j=1}^{M} \mathcal{CS}_j.
\end{eqnarray} 
Since non-Abelian gauge transformations are disallowed by Bloch Hamiltonian, $\mathcal{CS}_j$ and the Chern-Simons coefficient of ground state $\mathcal{CS}_{GS}$ are gauge-invariant.

The embedded $SU(2)$ matrix can be written in spin-$s$ representation, with $s=\frac{1}{2},1,..,\frac{N-1}{2}$. For spin $s$-representation, only $(2s+1)$ orbitals participate in Hopf map, and the Bloch Hamiltonian for this subspace assumes the form
\begin{equation} \label{general_Hamiltonian}
    \mathscr{H}_s(\bs{k}) = \sum_{i=1}^{2s} c_i [\bs{n}(\bs{k}) \cdot \bs{S}]^i.
\end{equation}
where $\bs{n}(\bs{k})$ is given by Eq.~\ref{model1}, $c_i$'s are \emph{constants that dictate band ordering}, and $(S_1, S_2, S_3 )$ are three generators of $\mathfrak{su}(2)$ algebra in spin-$s$ representation. The $SU(N)$ skyrmion number and $N_H$ are related by~\cite{carroll1993} \begin{eqnarray}
W[\mathscr{U},s] = \frac{2}{3} s (s+1) (2s+1) N_H. \end{eqnarray}
Only for the minimal $s=\frac{1}{2}$ embedding (essentially a decoupled two-band model), $W_{min}=W[\mathscr{U},s]=N_H$. The maximum skyrmion number $W_{max}=\frac{1}{6} N(N^2-1) N_H$ is realized for maximum spin $s=\frac{N-1}{2}$ embedding, \emph{for which all bands participate in Hopf map}. While bands participating in Hopf map support $\mathcal{CS}_j \propto N_H$, one cannot predict actual value of $\mathcal{CS}_j$ and $\mathcal{CS}_{GS}$ without the knowledge of spin-representation $s$, and $c_i$'s. For example, consider $s=\frac{3}{2}$ embedding for $4$-band model
\begin{equation}
 H\left(\bs{k}; N=4; s=\frac{3}{2}\right) =c_1 \bs{n}(\bs{k}) \cdot \bs{S} + c_2 (\bs{n}(\bs{k}) \cdot \bs{S})^2
 \end{equation}
with $c_3=0$. While $W[\mathscr{U}, s=\frac{3}{2} ]=10 N_H$ is independent of the ratio $c_2/c_1$, by tuning $c_2/c_1$ one can drive topological phase transition between two ground states with $\mathcal{CS}=5 N_H$ and $\mathcal{CS}=N_H$. Similar $N$-band models of HCIs can be constructed by using $\bs{n}(\bs{k})$ from Eq.~\ref{HCI}. Calculations of Witten effect for $N$-band HIs and HCIs will be reported in a future publication.

\underline{Conclusion}: By studying Witten effect for test magnetic monopole, we have identified the precise relationship between isotropic magneto-electric coefficient and Hopf invariant for 2-band Hopf and Hopf-Chern insulators. We also demonstrated the stability of $N$-band Hopf insulators against adiabatic flattening of band dispersions and provided concrete examples of tight-binding models. 

For realistic, time-reversal symmetry breaking materials, multiple $SU(2)$ subgroups of $SU(N)$ can carry three-dimensional winding numbers of opposite sign. Hence, the diagonalizing matrix of Hamiltonian does not carry a net winding number. But the constituent bands and the ground state can support quantized Chern-Simons coefficient. Such examples can be found in Ref.~\cite{tynermonopole}. 

Compared to Hopf insulators, the Hopf-Chern insulators do not require the absence of Chern numbers. The models are simpler to construct by twisting stacked Chern insulators. Thus, insulating helical magnets can serve as potential material candidates for Hopf-Chern insulators.

\acknowledgements{
This work was supported by the National Science Foundation MRSEC program (DMR-1720139) at the Materials Research Center of Northwestern University, and the start up funds of P. G. provided by the Northwestern University. A part of this work was performed at the Aspen Center for Physics, which is supported by National Science Foundation grant PHY-1607611.}

\bibliographystyle{apsrev4-1}
\bibliography{refer1.bib}

\end{document}